\documentclass[5p,twocolumn]{elsarticle}

\usepackage{graphicx}
\usepackage{dcolumn}
\usepackage[caption=false]{subfig}
\usepackage{bm}
\usepackage{amssymb}
\usepackage{amsmath}
\usepackage{amsthm}
\usepackage{color}
\usepackage{url}
\usepackage{float}
\usepackage{placeins}

\definecolor{blue}{rgb}{0.3,0.3,0.9}
\definecolor{green}{rgb}{0.2,0.7,0.2}
\DeclareMathOperator{\erf}{erf}

\begin{document}

\begin{frontmatter}


\title{Finite size effects on critical correlations in momentum space}

\author[inst1]{A.~Brofas\corref{cor1}}
\ead{thanosb@phys.uoa.gr}
\author[inst1]{F.~K.~Diakonos}
\ead{fdiakono@phys.uoa.gr}

\address[inst1]{Department of Physics, University of Athens, Athens GR-15784, Greece}

\cortext[cor1]{Corresponding author}

\begin{abstract}

The search for the QCD critical end point (CEP) is a major objective of contemporary heavy-ion physics, motivating the study of fluctuation observables that are sensitive to critical dynamics. In particular, baryon-number fluctuations provide a natural probe because the net-baryon density can serve as an effective order parameter in the vicinity of the CEP. Near criticality, long-range correlations and power-law scaling are expected to emerge in the real-space two-point function of the baryon density, yet the finite size and finite lifetime of the fireball created in heavy-ion collisions impose intrinsic cutoffs that regulate the growth of the correlation length. These finite-size constraints significantly modify the observable structure of fluctuations, especially in momentum space, where experiments perform measurements. In this work we present a theoretical analysis of the momentum-space two-point correlation function for a system of finite spatial extent. We show that finite size effects lead to an effective scaling exponent which coincides with that of the infinitely extended system only in a prescribed scaling region within the experimentally accessible momentum range.

\end{abstract}


\end{frontmatter}
A central objective of current heavy-ion collision experimental programs at RHIC (BNL) \cite{STAR} and SPS (CERN) \cite{SHINE} is to confirm (or not) the existence of the QCD critical end point (CEP) -- a remnant of the chiral symmetry restoration -- in the QCD phase diagram and to identify its position. However, the strong interacting matter created in these collisions occupies only a tiny spatial region, set by factors such as nuclear size and centrality of the collision. Due to this, theoretical studies must account for how experimental signals could be influenced by the finite volume of the system, providing an important consistency check between experiment and theory \cite{BFI, BFIS}. In this setting, any relevant observable must capture signatures of critical behavior through the particle spectra measured in the transient states created during high-energy ion collisions, where only hadrons are accessible. Consequently, these observables, suitable for CEP detection, depend solely on fluctuations in hadronic momenta and are related to the chiral transition’s order parameter \cite{Antoniou2006}. 

In this context, fluctuations of conserved charges, most notably the baryon number density $n_B(\mathbf{x})$, are especially valuable to investigate. This is due to the fact that critical QCD thermodynamics can be expressed using $n_B(\mathbf{x})$, which serves as an alternative but equivalent order parameter \cite{Antoniou2006, Stephanov, Stephanov_Hatta} to the chiral (dipion) condensate $\langle \bar{q} q \rangle$. This choice is particularly suitable because it corresponds to directly measurable properties of the system's freeze-out state. Thus, within this framework, the density - density correlator $\langle n_B(\mathbf{x})n_B(\mathbf{0})\rangle$ becomes a central object of investigation. 

Taking into account the symmetry of nuclei collisions, in which we are interested in the transverse plane, we can define the two-dimensional density $\rho(\mathbf{x}_{\perp})$, with 
$\mathbf{x}_{\perp}$ denoting the transverse configuration space and investigate the corresponding density-density (connected) correlation function $\langle \rho(\mathbf{x}_{\perp})\rho(\mathbf{0})\rangle_c$.

Ultimately, to connect the theoretical momentum space fluctuations with experimental measurements, it is necessary to search for the appearance of scaling behavior in suitable observables defined in transverse momentum space, specifically in the case of finite systems. The reason is twofold. On the one hand, thermal systems of large but finite size form fractal structure in configuration space as they approach the critical point in the phase diagram. This geometric feature arises from finite-size scaling and holds for length scales that are smaller than the correlation length $\xi$, $|\delta \mathbf{x}|\ll \xi$. Their self-similar structure is characterized by a 
fractal dimension $d_F = \frac{\delta d}{\delta + 1}$, where 
$\delta$ is the isotherm critical exponent and $d$ is the topological dimension \cite{Antoniou_Davis_Diakonos}. It is a goal of this paper to examine how this property influences the correlations in (transverse) momentum space, taking into account finite size effects. 

On the other hand, a promising tool for probing the critical behavior of such systems, encoded in critical exponents, is intermittency methods based on the second factorial moment, which is measurable \cite{Antoniou2006}. One can derive relations connecting the behavior of the second factorial moment and the scaling exponent of momentum space correlations, and thus measure the effects that the system's finite size imposes.

With these considerations in mind, we proceed to analyze the density–density correlations in momentum space. To maintain full generality, we work in $d$ spatial dimensions. The relevant quantity is the Fourier transform of the two-point correlation function $\langle \rho(\mathbf{x})\rho(\mathbf{0})\rangle_c$  \cite{Antoniou_Davis_Diakonos}. Based on the discussion above, we assume a general power law behavior for this correlator in configuration space of the form: $|\mathbf{x}|^{-a}$. 

Starting with the case of a system with infinite size and using an obvious notation, we define the desired quantity:

\begin{equation}
\mathcal{F}_{\infty}^d \left[|\mathbf{x}|^{-a} \right] (|\mathbf{k}|) = \int \mathrm{d}^d x \, |\mathbf{x}|^{-a} \, \mathrm{e}^{-2\pi i \mathbf{k}\cdot\mathbf{x}}
\label{eq:Statement}
\end{equation}

For $d>a$ this integral can be evaluated by performing the 
$(d-1)$-dimensional surface integration and subsequently employing the integral representation of the Bessel functions, which leads to the result:

\begin{equation}
\mathcal{F}_{\infty}^d  (|\mathbf{k}|) = \pi^{a-\frac{d}{2}}\frac{\Gamma\left(\frac{d-a}{2}\right)}{\Gamma\left(\frac{a}{2}\right)} \frac{1}{|\mathbf{k}|^{d-a}}
\label{eq:InfiniteDomain}
\end{equation}

An alternate route is to use an integral representation of the power-law itself, namely:

\begin{equation}
\frac{1}{r^a} = \frac{2\pi^{\frac{a}{2}}}{\Gamma\left(\frac{a}{2}\right)} \int_0^\infty \mathrm{d}x \, x^{a-1} \, \mathrm{e}^{-\pi r^2 x^2}
\label{eq:PowerLawIntRep}
\end{equation}

By substituting Eq.~(\ref{eq:PowerLawIntRep}) into Eq.~(\ref{eq:Statement}), we arrive at the following intermediate relation:

\begin{equation}
\mathcal{F}_{\infty}^d(|\mathbf{k}|) = \frac{2\pi^{\frac{a}{2}}}{\Gamma\left(\frac{a}{2}\right)} \int_0^\infty \mathrm{d}\xi \, \xi^{a-1} \int \mathrm{d}^d \mathbf{x} \, \mathrm{e}^{-2\pi i \mathbf{k}\cdot \mathbf{x} - \pi |\mathbf{x}|^2 \xi^2}
\label{eq:Intermediate}
\end{equation}

We adopt this method because the introduction of the parameter 
$\xi$, besides defining a characteristic momentum scale, produces a Gaussian regulator. This exponential factor smooths the boundary contributions of the finite system. Consequently, the $d$-dimensional integral factorizes into a product of 
$d$ one-dimensional integrals of the form:

\begin{equation}
g(\xi; k)= \int_{-\infty}^\infty \mathrm{d}x \, \mathrm{e}^{-2\pi i k x - \pi x^2 \xi^2} = \frac{1}{\xi} \mathrm{e}^{-\frac{\pi}{\xi^2} k^2}
\label{eq:IntermediateIntegral}
\end{equation}

\noindent
where $k$ and $x$ refer to one-dimensional cases. By substituting Eq.~(\ref{eq:IntermediateIntegral}) into Eq.~(\ref{eq:Intermediate}) and utilizing the representation of Eq.~(\ref{eq:PowerLawIntRep}), we once again recover the infinite domain power-law of the momentum measure as seen in Eq.~(\ref{eq:InfiniteDomain}).

We can now transition to the finite-domain case. This is achieved by focusing on the integral in 
Eq.~ (\ref{eq:IntermediateIntegral}) and restricting the domain to a linear extent of size $\delta$. The resulting integral in the finite domain is:

\begin{equation}
g(\xi; k) = \frac{1}{\xi} \mathrm{e}^{-\frac{\pi}{\xi^2} k^2} \Re\left(\erf\left(\frac{\sqrt{\pi}}{\xi}(\delta\xi^2+ik)\right)\right)
\label{eq:IntermediateIntegralFinite}
\end{equation}

In order to simplify this expression, we use the known expansion of the error function with complex argument, as given in relation 7.1.29 of \cite{AS}. In the $d$-dimensional case, the expression would involve a product of $d$ integrals of the form in Eq. (\ref{eq:IntermediateIntegralFinite}). However, since we are ultimately interested in the two-dimensional transverse-momentum space, we restrict ourselves to 
$d=2$. We retain only the leading term in the expansion of the product of the real parts, namely $\erf^2(\sqrt{\pi} \delta \xi)$. This is justified by the assumption that the linear extent is small, $\delta \ll 1$. Under these conditions, denoting $k=|\mathbf{k}|$ and changing variables from 
$\xi$ to $x^{-1}$, we obtain:

\begin{equation}
\mathcal{F}_{\delta}^{d=2}(k) = \frac{2\pi^{\frac{a}{2}}}{\Gamma\left(\frac{a}{2}\right)} \int_0^\infty \mathrm{d}x \, x^{1-a} \, \mathrm{e}^{-\pi k^2 x^2} \, \erf^2(\sqrt{\pi} \delta x)
\label{eq:StatementFiniteVer1}
\end{equation}

A more useful form emerges by employing the definition of the squared error function in the two-dimensional domain and transforming to polar coordinates. This yields:

\begin{equation}
\erf^2(s) = 1-\frac{4}{\pi}\int_0^1 \mathrm{d}t \, \frac{\mathrm{e}^{-s^2(1+t^2)}}{1+t^2}
\label{eq:ErrorFuncPolar}
\end{equation}

By substitution to Eq.~(\ref{eq:StatementFiniteVer1}) we obtain two terms. The first term is the infinite domain case of Eq.~(\ref{eq:InfiniteDomain}) for $d=2$. The second term is the correction $I$ due to the finite size of the system: 

\begin{equation}
\mathcal{F}_{\delta}^{d=2}(k) = \mathcal{F}_{\infty}^{d=2}(k) - I(\delta ; k)
\label{eq:InfiniteMinusCorrection}
\end{equation}

We are interested in the behavior of the correction term. The correction $I(\delta ; k)$ contains an integral that appears in the argument of the squared error function, which is recognized as the integral representation of the modified Bessel function of the second kind $\mathcal{K}$. Therefore, we can rewrite the correction in the more useful form:

\begin{equation}
\begin{split}
I(\delta ; k) = & \frac{8 \pi^{\frac{a}{2}-1}}{\Gamma\left( \frac{a}{2} \right)} \left( \frac{\delta}{k} \right)^{1-\frac{a}{2}} \times \\
&\int_0^1 \mathrm{d} t \, (1+t^2)^{-\frac{a+2}{4}} \, \mathcal{K}_{1-\frac{a}{2}}(z \sqrt{1+t^2}),
\end{split}
\label{eq:CorrectionTerm}
\end{equation}

\noindent 
where $z = 2\pi k\delta$. We are now ready to investigate the two regimes of interest, namely: $k\delta \gg 1$ or $z\rightarrow \infty$ and $k\delta \ll 1$ or $z \rightarrow 0$.

In the regime $k\delta \gg 1$, we use the asymptotic form of the modified Bessel function of the second kind. Since the integrand has a global minimum, we can apply Laplace’s method. After evaluating the resulting Gaussian integrals, we obtain the correction in the large-$k \delta$ limit:

\begin{equation}
I_{\gg} = 4 \frac{\pi^{\frac{a}{2}}}{\Gamma\left( \frac{a}{2} \right)}  \left( \frac{\delta}{k} \right)^{1-\frac{a}{2}} \frac{\mathrm{e}^{-z}}{z} \cdot\left( 1 - \frac{2a+3}{8z} +\mathcal{O}\left(\frac{1}{z^2}\right)\right)
\label{eq:CorrectionLarge}
\end{equation}

To leading order, we can safely say that the correlations in question, as expressed by $\mathcal{F}_{\delta}^{d=2}$, are dominated by the infinite-domain behavior as expressed by $\mathcal{F}_{\infty}^{d=2}$. Therefore, at large $k\delta \gg 1$, in two dimensions, the correlations exhibit the conventional power law $k^{-(2-a)}$.

We can now continue to the regime defined by: $k\delta \ll 1$ or $z \rightarrow 0$. For this regime, we again use the approximation of the modified Bessel function of the second kind for small arguments. In this process, we discover that the first order correction to the integral in Eq.~(\ref{eq:CorrectionTerm}) is identical to the infinite domain term and so the terms cancel due to the minus sign. In order to obtain non-trivial correlations in this regime, we must use higher order terms. In second order, we acquire a term that is independent of the momentum magnitude $k$. In third order, we obtain the first 
non-canceling, non-trivial $k$-dependent behavior. We summarize these results below:

\begin{equation}
I_{\ll} = I^{(1)} +I^{(2)}+I^{(3)}
\label{eq:CorrectionSmall}
\end{equation}

\begin{equation}
\begin{split}
    I^{(1)} & = \mathcal{F}_{\infty}^{d=2} \\
    I^{(2)} & = 4 \frac{\Gamma \left( \frac{a}{2}-1 \right)}{\Gamma \left( \frac{a}{2} \right)} S(a) \,\delta^{2-a} = C \\
    I^{(3)} & = 8 \frac{\pi^a}{a}\frac{\Gamma \left( 1-\frac{a}{2} \right)}{\Gamma \left( \frac{a}{2} \right)} \, \delta^2 \, k^a = D \, k^a,
\end{split}
    \label{eq:CorrectionSmallDetails}
\end{equation}

\noindent 
where $S(a) = \int_0^1 \mathrm{d}t \, (1+t^2)^{-a/2}$. Thus, the correlations in this regime are determined by: 

\begin{equation}
\mathcal{F}_{\delta}^{d=2} = -C -D \, k^a
\label{eq:CorrelationsSmall}
\end{equation}

With the small and large $k \delta$ regime expressions, we can now examine the full behavior of the correlations with respect to both the real space extent of the system governed by $\delta$ and the values of the momenta governed by $k$. Assuming we have estimates for the extent of correlations in real space, which will be analyzed later, we can fix $\delta$ and examine the system with respect to the momenta $k$. It is obvious that for very small $k$, the behavior is dominated by the constant $C$ and so the correlations assume a constant value and the power law disappears. For very large $k$ the $I_{\gg}$ rapidly diminishes and the behavior is governed by the conventional, infinite domain, power law of $k^{-(2-a)}$. It is natural to infer that there is an intermediate crossover window $k_{\rm left} \leq k \leq k_{\rm right}$ in which the transition between the two regimes takes place. It is our next topic of interest to define this window.

For the left hand boundary, we note first that for values of $a \in (0,2)$, the constant $C$ is negative. As a subtle note in passing, additional terms appear in the small argument expansion of the modified Bessel function of the second kind, of the form $\mathrm{ln}(k\delta)$ when its order is integer. Since we wish to focus in non-integer $a$ and therefore non-integer Bessel function order, we do not include such terms. With these considerations, the left boundary is obtained when the small non-constant in $k$ correction equals the constant term. One could possibly variate this boundary slightly by tuning the ratio between the two terms so that it differs from unity. Thus:

\begin{equation}
k_{\rm left} = \left( \frac{|C|}{D}\right)^{1/a}
\label{eq:kleft}
\end{equation}

At very large $k$, the behavior is dominated by $\mathcal{F}_{\infty}^{d=2}$, while at lower $k$ the term $I_{\gg}$ competes. At the crossover region, we should find the $I_{\gg}$ term being of magnitude comparable to that corresponding to the infinitely extended system. We then define this boundary $k_{\rm right}$ by the equation: $I_{\gg}(k_{\rm right}) = \mathcal{F}^{d=2}_{\infty}(k_{\rm right})$ which results in the transcendental equation:

\begin{equation}
k_{\rm right}^{-\frac{a}{2}} \mathrm{e}^{-2\pi k_{\rm right} \delta} = \frac{(\pi \delta)^{\frac{a}{2}}}{2} \Gamma \left( 1- \frac{a}{2}\right)
\label{eq:transendental}
\end{equation}

Within the crossover window defined by $(k_{\rm left},k_{\rm right})$ we are to investigate any power law behavior that could result to an effective exponent. Inside this region, before the term $I_{\gg}$ becomes sufficiently large, the two term expression of Eq.~(\ref{eq:CorrelationsSmall}) is a good approximation. The power law that emerges in this region (depending on $k$) is characterized by the exponent $\gamma_{\rm eff}$:

\begin{equation}
\gamma_{\rm eff} = \frac{\mathrm{d \,ln}\mathcal{F}_{\delta}^{d=2}}{\mathrm{d \, ln}k},
\label{eq:gammaeff}
\end{equation}

\noindent 
where we note in passing that if $a$ is not within $(0,2)$, the absolute value of $\mathcal{F}_{\delta}^{d=2}$ should be employed.

To further explore the implications of these results, we perform numerical calculations. In short, the algorithm computes the coefficients $C$ and $D$ and performs a numerical integration of the correction term $I$ of Eq.~(\ref{eq:CorrectionTerm}). Then it computes the values for $\mathcal{F}_{\delta}^{d=2}$ and $\mathcal{F}_{\infty}^{d=2}$ and finally the exponent $\gamma_{\rm eff}$. It also computes the boundary $k_{\rm left}$ using Eq.~(\ref{eq:kleft}), while regarding $k_{\rm right}$, for simplicity, instead of solving the transcendental equation Eq.~(\ref{eq:transendental}) it simply finds the value of $k$ for which $\gamma_{\rm eff}$ is within 0.01 of the asymptotic slope $-(2-a)$. Lastly, it produces a representative average value for the exponent within the crossover window, in order to demonstrate the change from the asymptotic value.

As for the parameters of the computation, first, concerning the exponent $a$ we use the the value $1/3$, referring to the critical exponent corresponding to particle density geometry induced by the 3D Ising model universality class \cite{Antoniou2006}. This is a widely accepted conjecture for the universality class of the CEP of the QCD phase diagram \cite{Wilczek}. For the linear extent $\delta$ we choose the value of $1$ measured in $\rm fm$. This is justified by the following reasoning: The distance $\delta$ should not be interpreted as the extent of the whole system, but really the extent of correlated baryons. A larger distance, for example of $7 ~ \rm fm$, comparable to the radius of nuclei used in heavy ion collisions, would determine the geometric extent of the system. Finite size effects however correspond to the spatial scale over which baryons remain correlated, which close to criticality can reach the extent of the whole system. In such case of finite system, critical power laws are replaced by finite sized modified scaling forms \cite{Ichihara}. It is safe to assume an extent of correlations of the order of $1$ to $10 ~ \rm fm$ \cite{Ling, Berdnikov, Stephanov2, Athanasiou}. 

Our theoretical and numerical computations dictate the emergence of three distinct regions with respect to the momenta. The first is the one that any power law behavior that is expressed as an exponent, vanishes. This is to be expected as the finite system imposes a maximum correlation length $\xi \leq \delta$. Thus fluctuations on length scales larger than $\delta$ simply do not exist and so momentum modes probing wavelengths of $k^{-1} \gg \delta$ cannot resolve any further structure. Hence we can only recover the total integrated strength of correlations inside the finite domain. This is connected to the baryon number susceptibility $\chi_B$ as:

\begin{equation}\label{eq:Susceptibility}
    \mathcal{F}_{\delta}^{d=2}(|\mathbf{k}| \ll 1/\delta) \overset{\mathrm{e}^{i\mathbf{k}\cdot\mathbf{x} \approx1}} {\approx} \int_{|\mathbf{x}| \lesssim \delta } \mathrm{d^2} \mathbf{x} \, \langle \rho(\mathbf{x}) \rho(\mathbf{0})\rangle_c  =  \chi_{B,\delta}
\end{equation}

\begin{figure}[H]
  \centering

  \subfloat[]{%
    \includegraphics[width=\columnwidth]{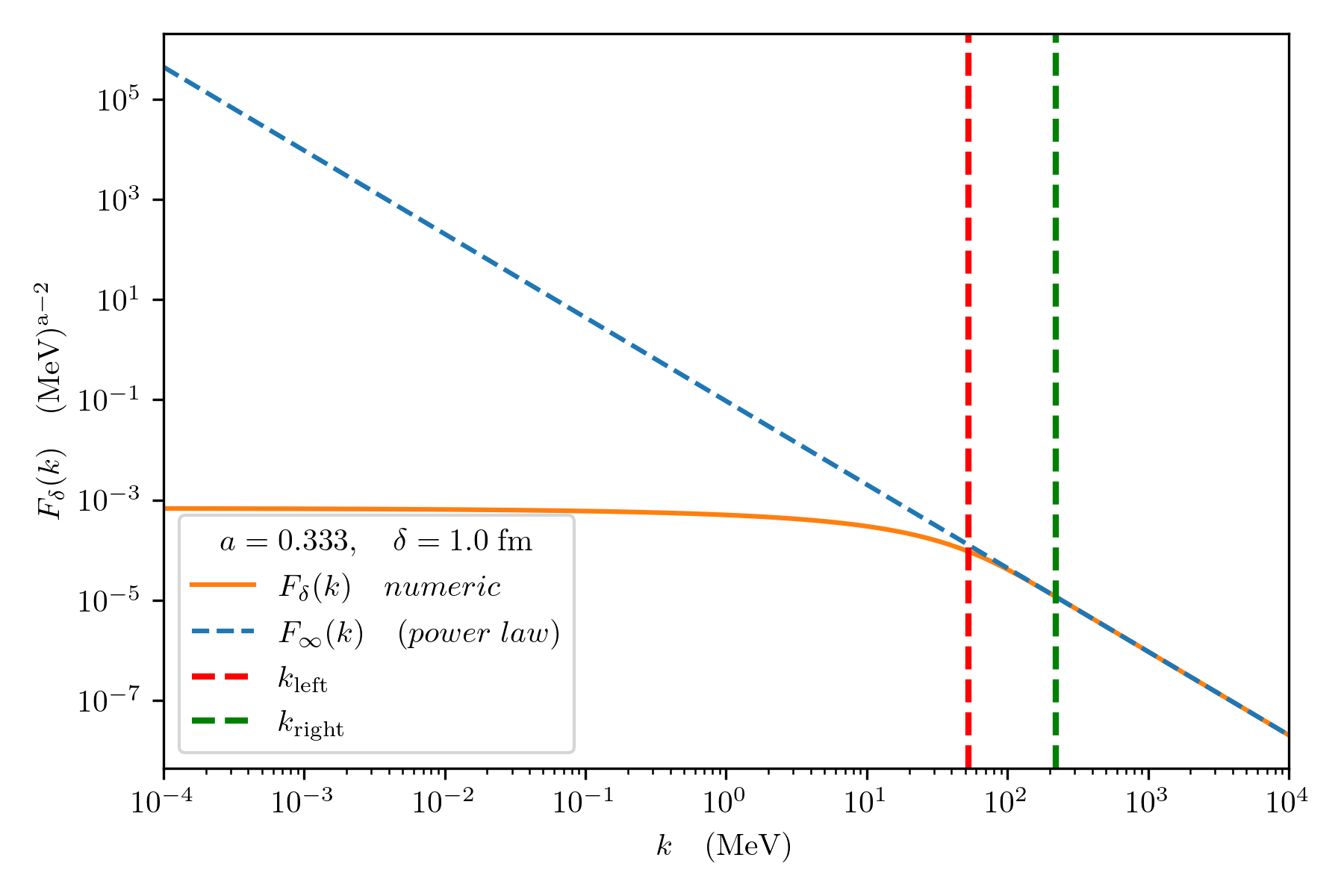}%
    \label{fig:fig1a}
  }

  \vspace{0.6ex}

  \subfloat[]{%
    \includegraphics[width=\columnwidth]{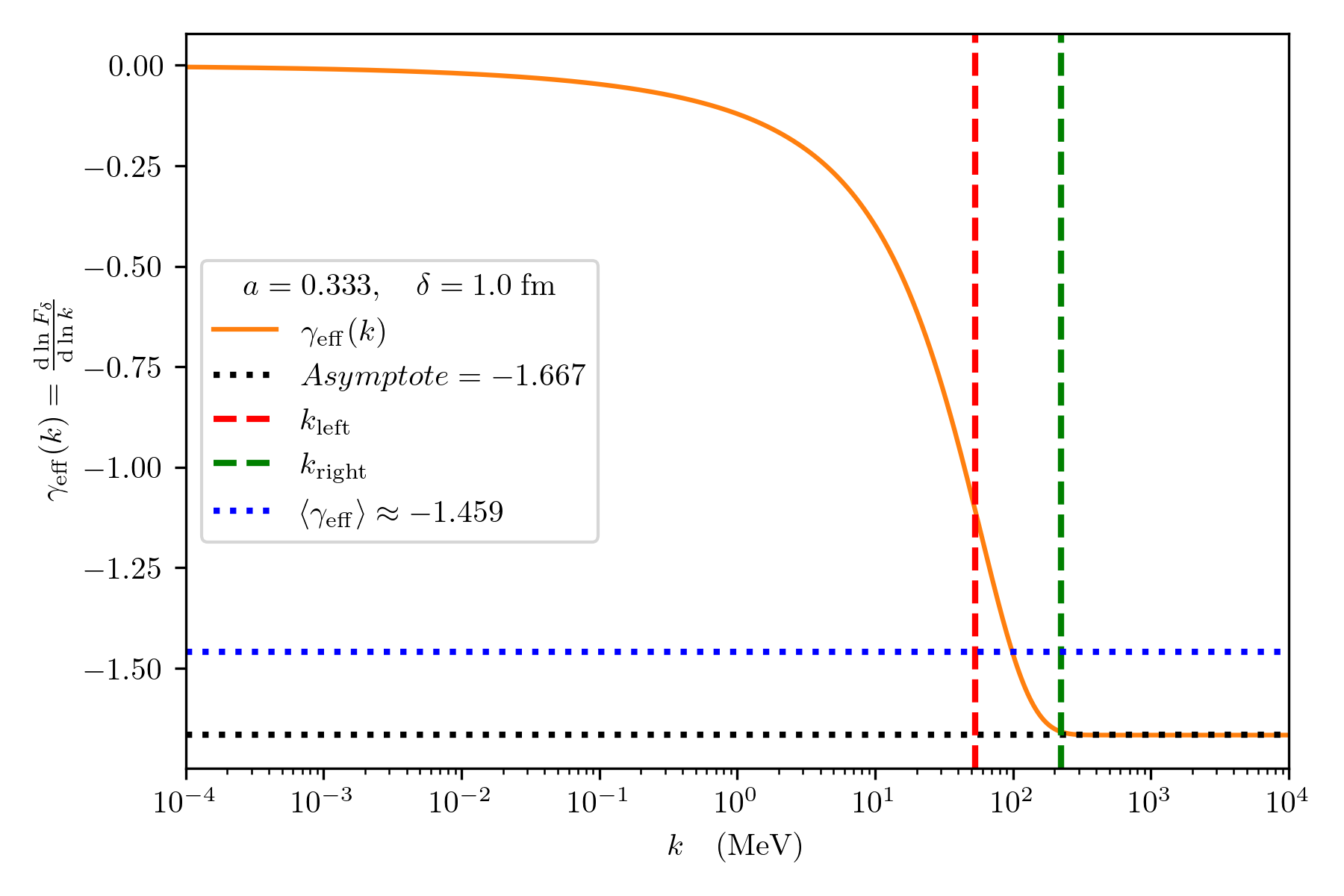}%
    \label{fig:fig1b}
  }

    \caption{Plots for fixed value of $\delta = 1 ~ \rm fm$ and $a=1/3$. (a) The momentum space correlations $\mathcal{F}_{\delta}^{d=2}$ as a function of the measure of the two dimensional momentum $k$. The blue dashed line is the infinite domain power law. While the vertical dashed lines are the boundaries $k_{\rm left}$, $k_{\rm right}$. (b) The exponent of $\gamma_{\rm eff}$ for different values of $k$ as derived from $\mathcal{F}_{\delta}^{d=2}$. The blue dashed line is the representative value of the exponent (average) while the black dashed line is the asymptote. Again the vertical lines are the boundaries.}     
\label{fig:fig1}
\end{figure}

Using the power law correlator one recovers the $\delta^{2-a}$ behavior that is predicted by Eq.~(\ref{eq:CorrectionSmallDetails}). Hence the constant plateau that is observed in Fig.~\ref{fig:fig1} (more pronounced in (a)) is due to this saturation effect. Another interesting feature of this region, that is connected to intermittency and experimental considerations, is the fact that since the susceptibility behaves as $\delta^{2-a}$, we can relate it to the size of nuclei, which in turn is linked to the freeze-out temperature. Consequently, these susceptibility values are connected to the freeze-out temperature. By examining this relationship, and in particular the emergence of a maximum or plateau with respect to the freeze-out temperature, one can approximate the critical temperature and identify an additional signature of the system's criticality. In addition, this saturation region sheds new light upon intermittency analysis based on the second factorial moment, since for small bin sizes, and equivalently for small momenta the second factorial moment saturates to a constant value at a sufficiently large bin number, when dynamical correlations are either absent or suppressed \cite{STAR, Wu, Wu2}.

The second region (between the dashed vertical lines in Fig.~\ref{fig:fig1} (a,b) contains the crossover window which lies in the momentum region of the order of $10^2$ MeV for $\delta \sim 1 \rm~ fm$. The last region is the one in which we recover the 
$k^{-(2-a)}$ - behavior, characteristic for the infinite system. This is because here, the momentum modes probe distances smaller than the cutoff $\delta$ and so boundary effects become irrelevant. In essence, the correlator behaves locally as if the system were infinite.

\begin{figure}[H]
  \centering

\subfloat[]{%
    \includegraphics[width=\columnwidth]{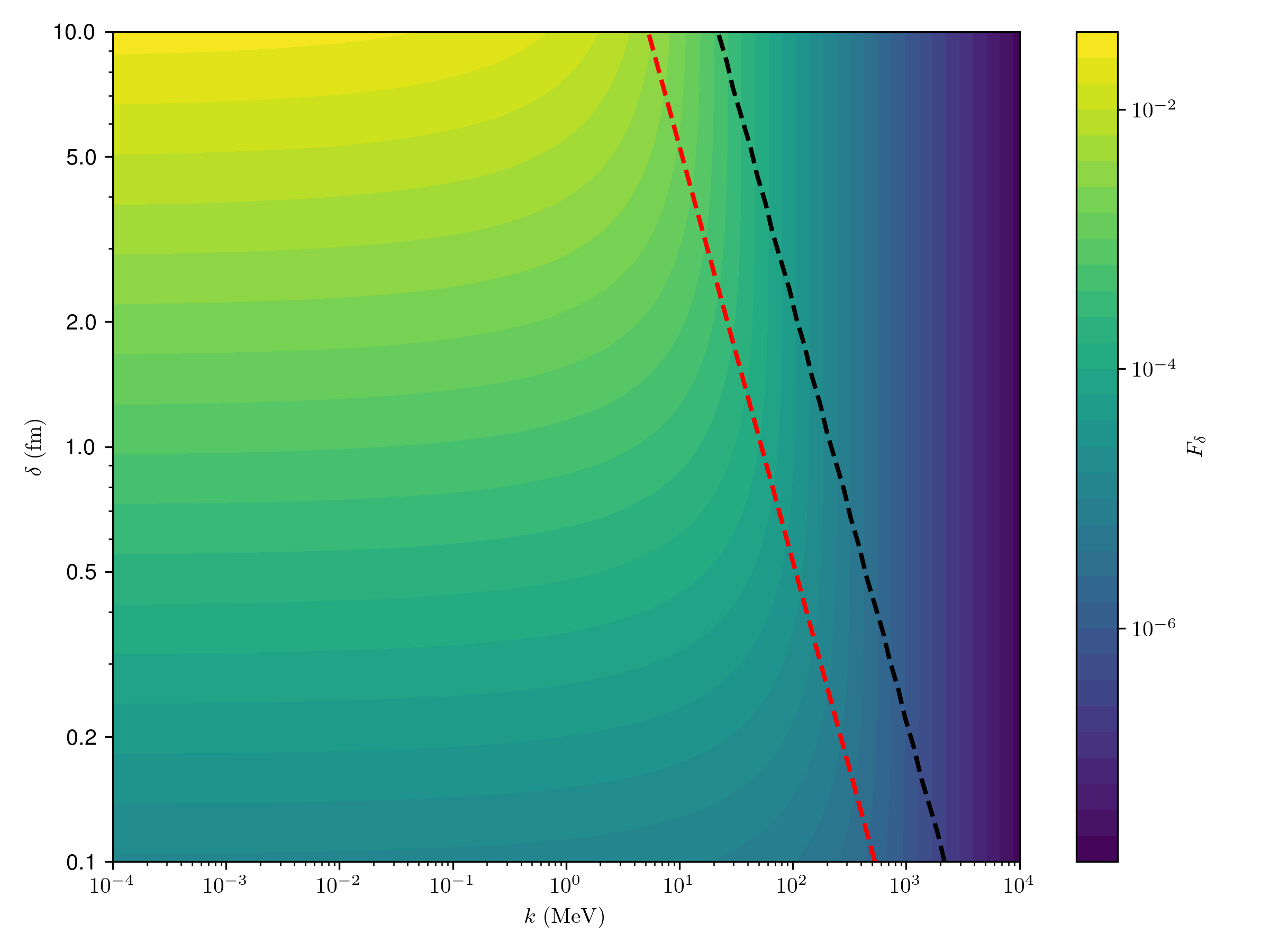}%
    \label{fig:fig2a}
  }

  \vspace{0.6ex}

  \subfloat[]{%
    \includegraphics[width=\columnwidth]{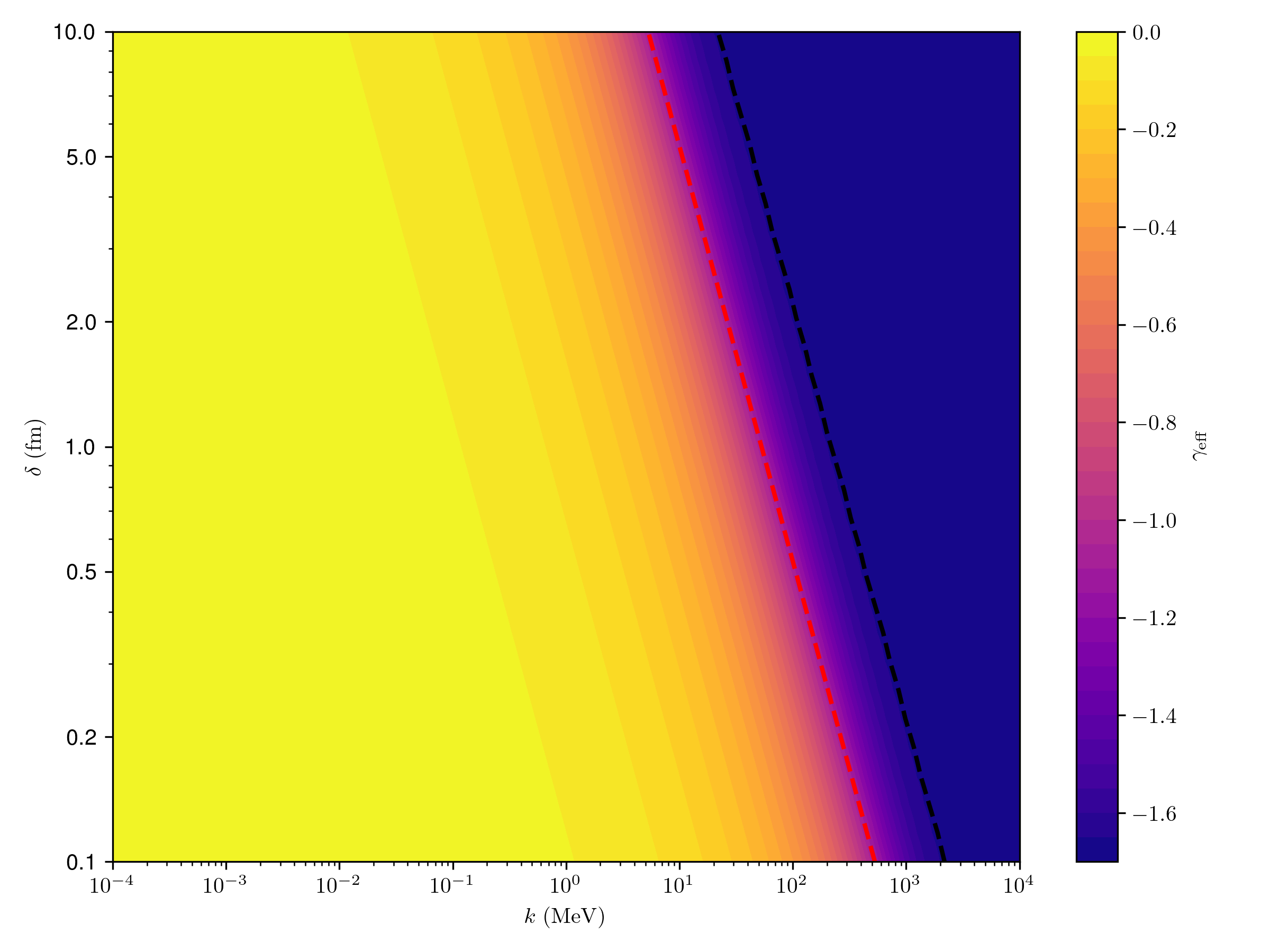}%
    \label{fig:fig2b}
  }

    \caption{Contour plots for varying values of $\delta$ and $a=1/3$. (a) The momentum space correlations $\mathcal{F}_{\delta}^{d=2}$ as a function of the measure of the two dimensional momentum $k$ and $\delta$. The dashed lines are the boundaries $k_{\rm left}$, $k_{\rm right}$. (b) The exponent $\gamma_{\rm eff}$ for different values of $k$ and $\delta$ as derived from $\mathcal{F}_{\delta}^{d=2}$. Again the dashed lines are the boundaries.}     
\label{fig:fig2}
\end{figure}

Finally, it is evident, that for different $\delta$, that is, system size (or more specifically maximum extend of correlations), we obtain a shifted window of momenta for the crossover region. The larger the system size, or equivalently the heavier the nuclei colliding in a heavy ion collision, the more the window is shifted to smaller momenta. These results are portrayed in Fig.~\ref{fig:fig2}.


In order to generalize our theoretical analysis and model more closely the physical system, we proceed and include a hard-core interaction for the protons. This translates to a vanishing of correlations in real space, below a threshold $r_0$. In order to differentiate from the previous case, we denote the upper extent of our system with $r$. By performing an analogous analysis with the one described in the previous cases, we arrive at the following expression for the correlations in momentum space for two dimensions:
\begin{equation}
\mathcal{F}^{d=2}_{r,r_0} = I(r,\mathbf{k};r_0/r) + I(r_0,\mathbf{k};r/r_0) - I(r,\mathbf{k};1) - I(r_0,\mathbf{k};1)
\label{eq:F_HC}    
\end{equation}
where we denote with $I(x,\mathbf{k};\Lambda)$ the generalized correction term:
\begin{equation}
\begin{split}
I(x,\mathbf{k};\Lambda) = & 8 \frac{\pi^{\frac{a}{2}-1}}{\Gamma \left( \frac{a}{2}\right)} \left(\frac{x}{|\mathbf{k}|}\right)^{1-\frac{a}{2}} \times \\
& \times \int_0^{\Lambda} \mathrm{d}t \, (1+t^2)^{-\left(\frac{a+2}{4}\right)}\mathcal{K}_{1-\frac{a}{2}} (2\pi |\mathbf{k}| x \sqrt{1+t^2})
\end{split}
\label{eq:Generalized_correction_term}    
\end{equation}

With these expressions in hand, first and foremost we are able to verify the reducibility of $\mathcal{F}^{d=2}_{r,r_0}$ in the case of $r_0 \rightarrow 0$ as a consistency check. Going, forward in a similar manner as in the conventional case, we are able to find the asymptotic behavior of the correlations in momentum space for small and large values of the magnitude of the momenta. Specifically, we find:

\begin{equation}
\begin{split}
    \mathcal{F}_{r,r_0}^{d=2} \overset{|\mathbf{k}| \gg 1}{\rightarrow} & 0 \\
    \mathcal{F}_{r,r_0}^{d=2}  \overset{|\mathbf{k}| \ll 1}{\rightarrow} &4\frac{\Gamma\left(\frac{a}{2}-1\right)}{\Gamma\left(\frac{a}{2}\right)} \times \\
    &\left\{ r^{2-a} \Delta S\left(a;\frac{r_0}{r} \right) + r_0^{2-a}  \Delta S\left(a;\frac{r}{r_0} \right)\right\} \\
    & -8 \frac{\pi^a}{a} \frac{\Gamma\left(1-\frac{a}{2}\right)}{\Gamma\left(\frac{a}{2}\right)}(r-r_0)^2 |\mathbf{k}|^a  \\
    \text{where:} & \\
    \Delta S(a;\Lambda)  = &S(a;\Lambda)-S(a;1) \\
    S(a;\Lambda)  = &\int_0^{\Lambda} \mathrm{d}t \, (1+t^2)^{-\frac{a}{2}}
\end{split}
    \label{eq:Correlations_HC_asymptotics}
\end{equation}

\begin{figure}[H]
  \centering

  \subfloat[]{%
    \includegraphics[width=\columnwidth]{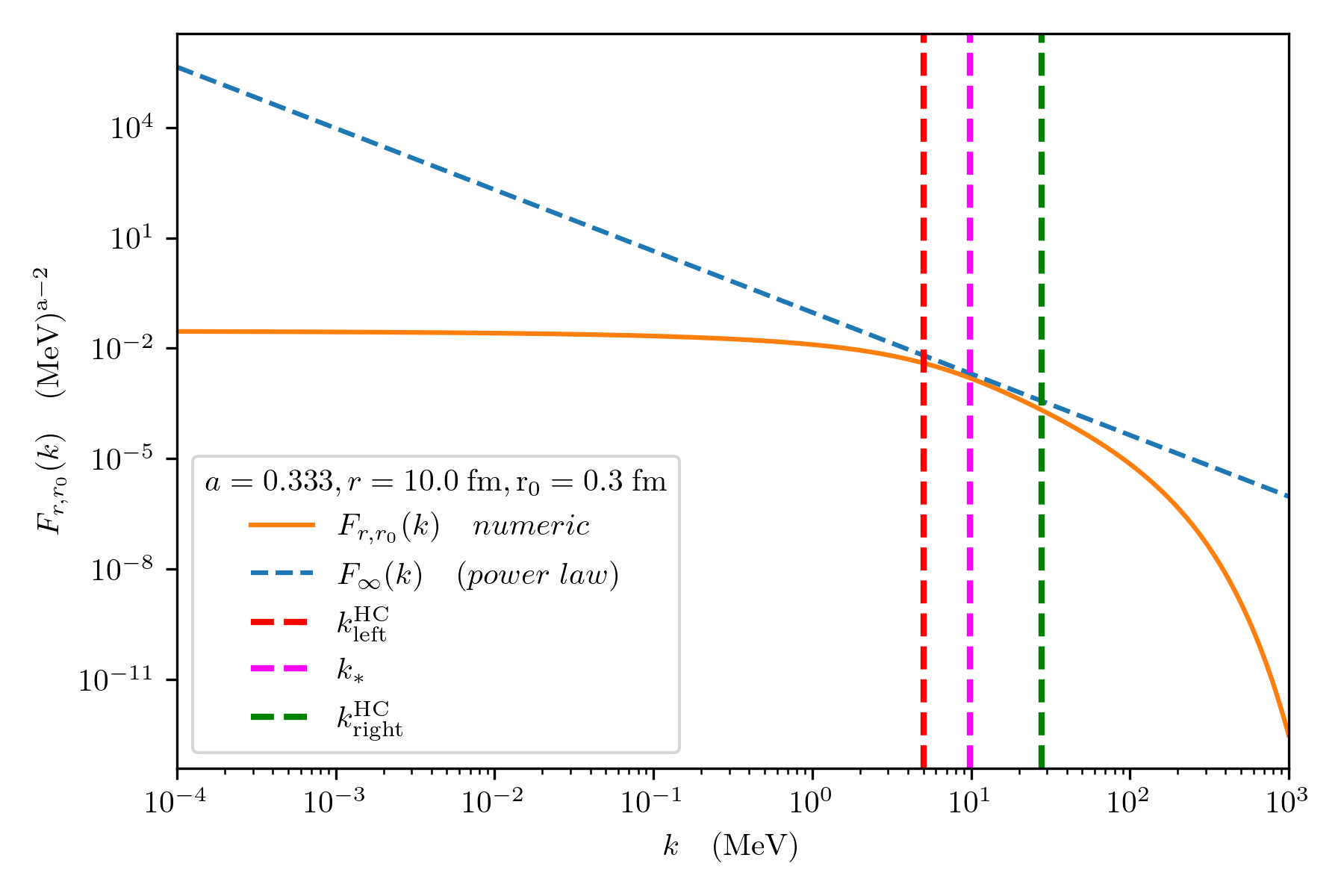}%
    \label{fig:fig3a}
  }

  \vspace{0.6ex}

  \subfloat[]{%
    \includegraphics[width=\columnwidth]{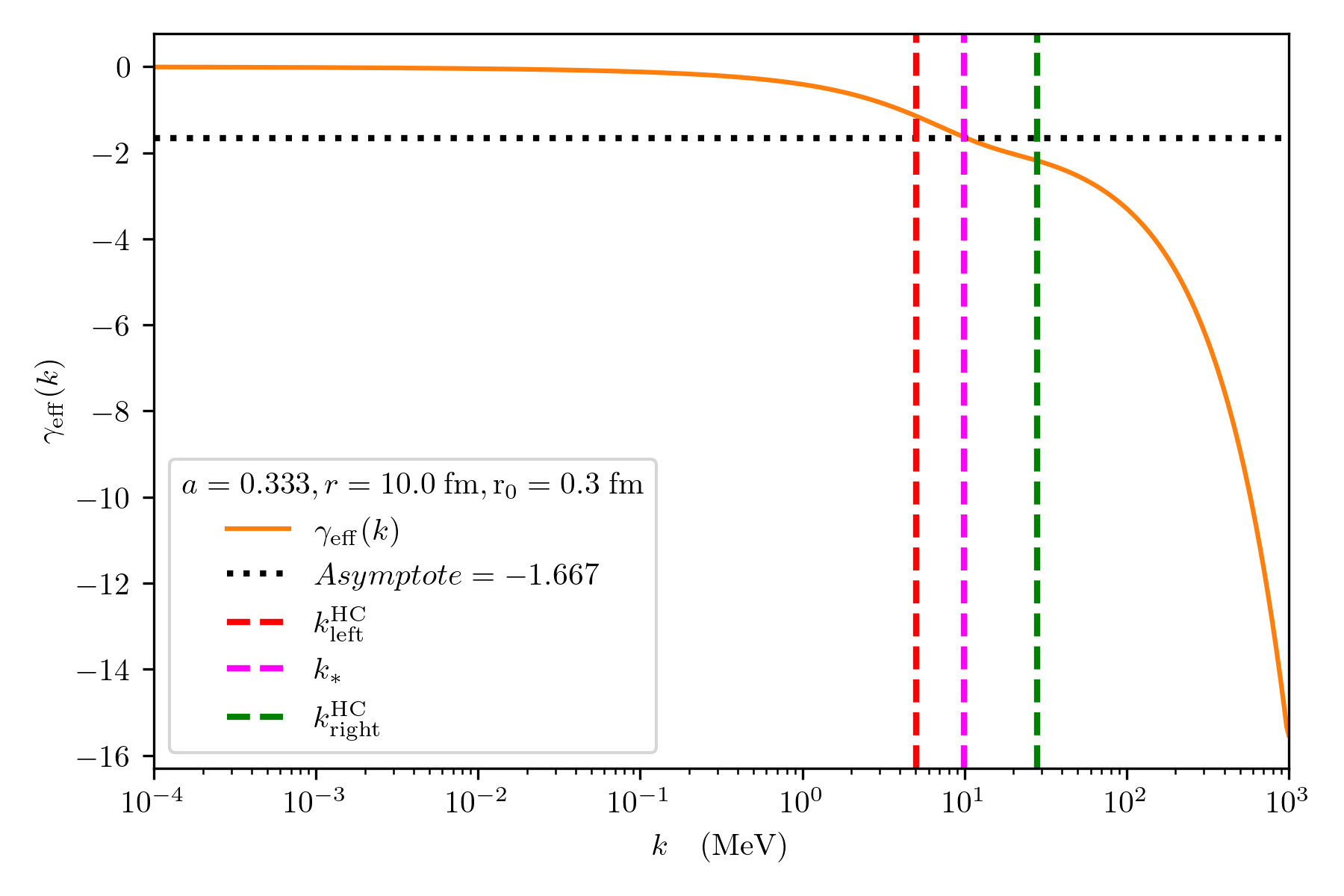}%
    \label{fig:fig3b}
  }

    \caption{Plots for fixed value of $r = 10 ~ \rm fm$, $r_0= 0.3 ~ \rm fm$ and $a=1/3$. (a) The momentum space correlations $\mathcal{F}_{r,r_0}^{d=2}$ as a function of the measure of the two dimensional momentum $k$. The blue dotted line represents the infinite domain power law behavior while the dotted vertical red and green lines the left and right boundaries respectively. The magenta dotted line represents the value of $k$ for which the $-(2-a)$ value for the exponent is attained. (b) The exponent of $\gamma_{\rm eff}$ for different values of $k$ as derived from $\mathcal{F}_{r,r_0}^{d=2}$. The dotted vertical red and green lines the left and right boundaries respectively. The magenta dotted line represents the value of $k$ for which the $-(2-a)$ value for the exponent is attained.}     
\label{fig:fig3}
\end{figure}

\begin{figure}[H]
  \centering

  \subfloat[]{%
    \includegraphics[width=\columnwidth]{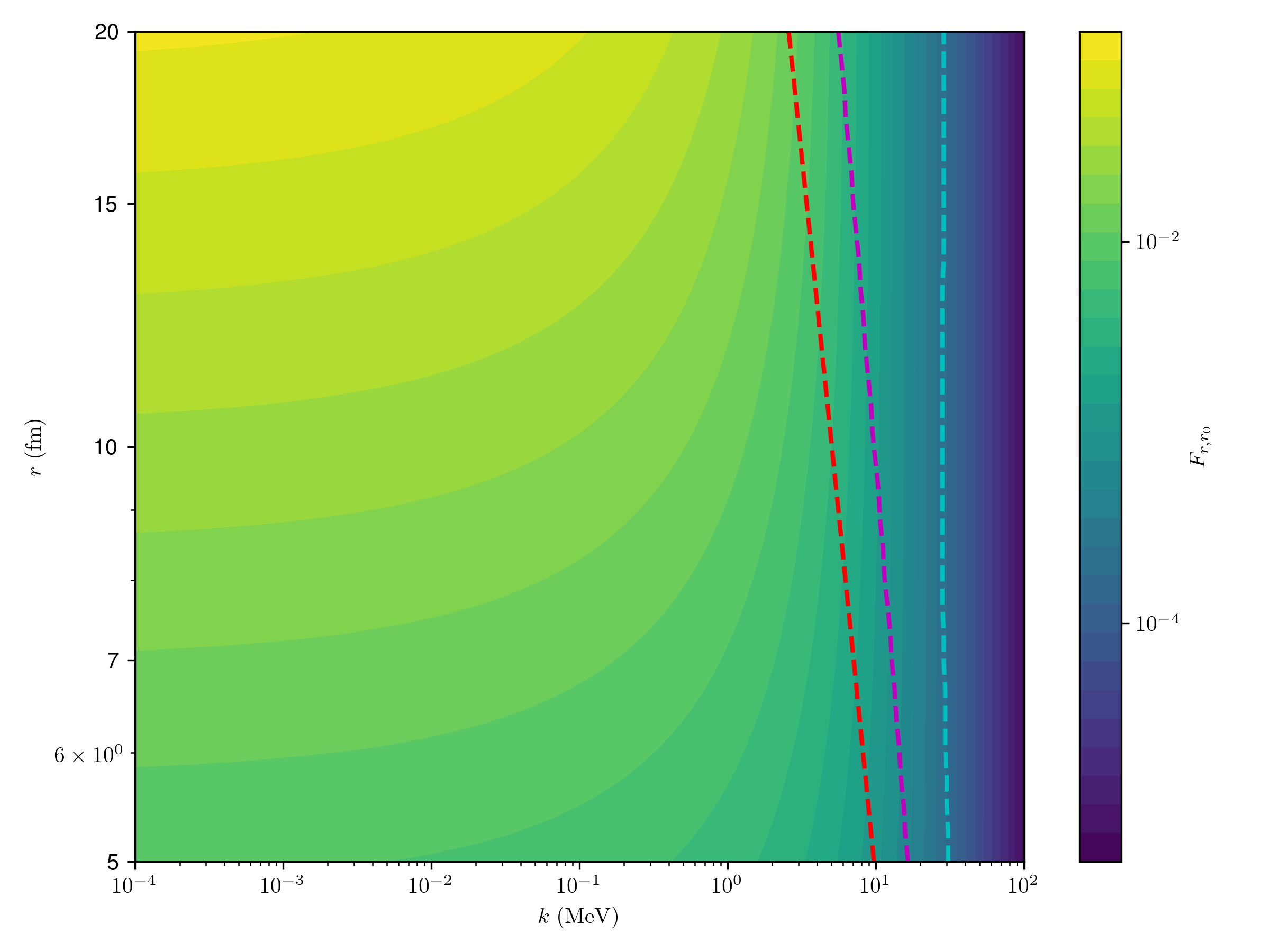}%
    \label{fig:fig4a}
  }

  \vspace{0.6ex}

  \subfloat[]{%
    \includegraphics[width=\columnwidth]{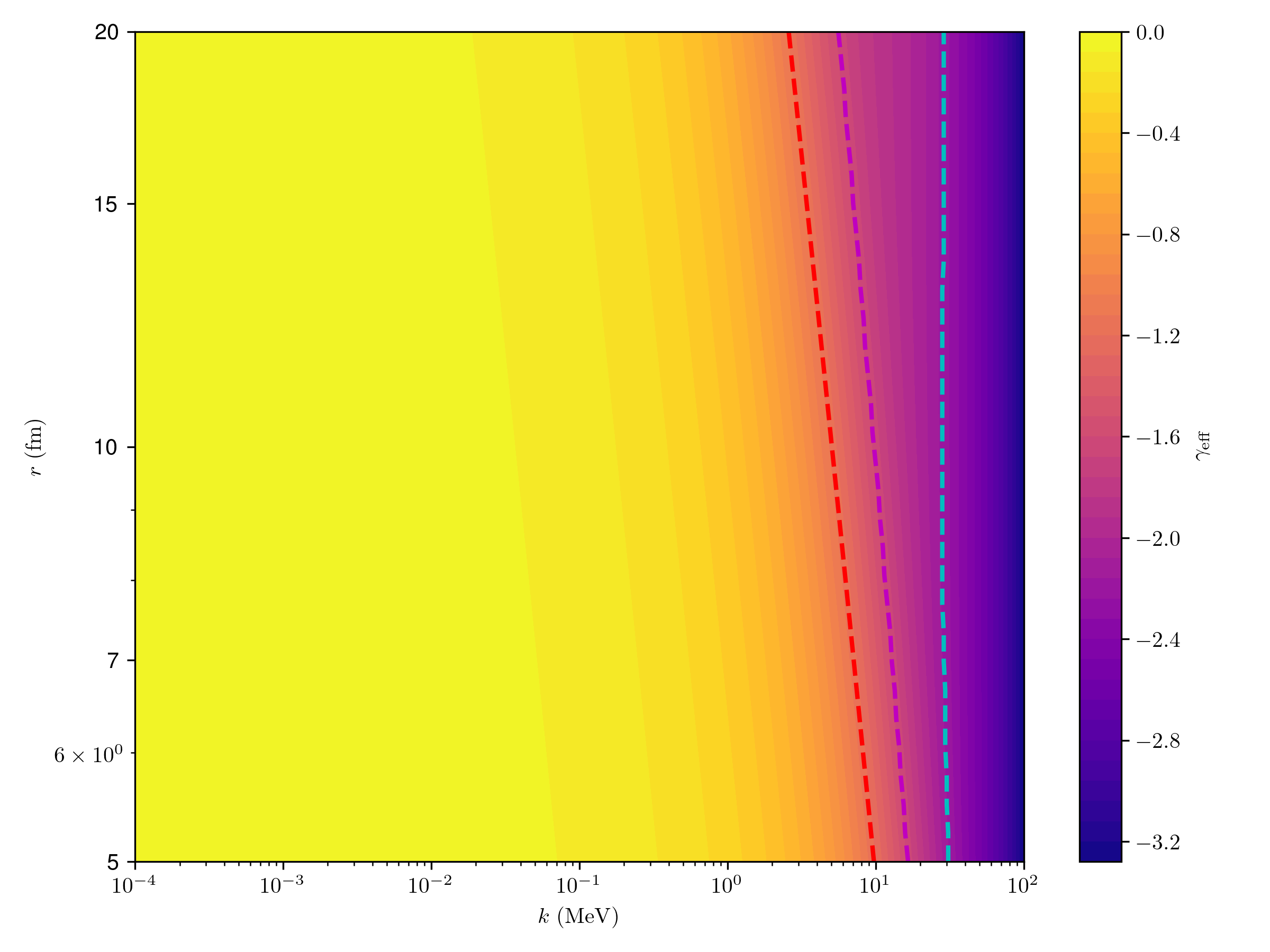}%
    \label{fig:fig4b}
  }

    \caption{Contour plots for varying values of $r$ and $a=1/3$. (a) The momentum space correlations $\mathcal{F}_{r,r_0}^{d=2}$ as a function of the measure of the two dimensional momentum $k$ and $r$. The red and cyan dotted lines represent the left and right boundaries respectively while the magenta one represents the value of $k$ for which the $-(2-a)$ value for the exponent is attained. (b) The exponent $\gamma_{\rm eff}$ for different values of $k$ and $r$ as derived from $\mathcal{F}_{r,r_0}^{d=2}$. The red and cyan dotted lines represent the left and right boundaries respectively while the magenta one represents the value of $k$ for which the $-(2-a)$ value for the exponent is attained.}     
\label{fig:fig4}
\end{figure}

From the behavior displayed in the above expressions, we deduce the following: For small momenta, or in the IR regime, the behavior of the system is merely shifted by the hole that is created by excluding an area of linear extent $r_0$ around zero. We still obtain a saturation effect with a shifted saturation constant, since we are again probing outside the system's limits. Contrary to the conventional case, in the large momenta, or in the UV regime, we notice that correlations die off. This is to be expected, since now there is no system below $r_0$ and so for $|\mathbf{k}| \gg 1/r_0$ we are probing inside the excluded region. In the intermediate region defined by $1/r \lesssim k \lesssim 1/r_0$, we cross over from the constant value of correlations to their exponential drop off. It is within this region that we briefly attain an exponent whose value corresponds to that of the ideal, infinitely extended system. Finally, as we expect, the larger the ratio $r/r_0$ is, the wider the window within which we obtain the infinite domain's power law will be. For the numerical investigation, we performed similar calculations. We calculate $k^{\rm HC}_{\rm left}$ from the coefficients of the small momentum expansion as before, as a signal of the departure from the saturation regime. For $k^{\rm HC}_{\rm right}$, we calculate it as follows: If the value of the exponent for $k^{\rm HC}_{\rm left}$ is $\Delta\gamma$ away from the ideal value of $-(2-a)$ then $k^{\rm HC}_{\rm right}$ is defined as the value of momenta for which the value of the exponent is $-\Delta\gamma$ away from the ideal value. In addition, we calculate $k_{*}$ which corresponds to the value of momenta for which the exponent attains the infinite domain value of $-(2-a)$. Finally, according to literature, we set the hard-core distance below which correlations vanish to be of the order of $r_0 \sim 0.3 ~\rm fm$ \cite{Rischke, Andronic, Braun}. For these results we refer to Fig.~\ref{fig:fig3} and Fig.~\ref{fig:fig4}, as well as to table~\ref{tab:window_values}.

\section*{Conclusions}

In this paper, we investigated how critical correlations of baryons (protons) in momentum space are modified when the system is restricted to a small but finite, real space, extent. Initially, we solved the problem without assuming a minimum distance corresponding to a hard-core interaction, that is a minimum non zero distance that the baryons can approach. We found that in the IR regime, the correlations are constant, reflecting the fact that in this regime we are probing large distances beyond the limits of the system, and we obtain the integrated correlations, connected to the susceptibility as seen in Eq.~\ref{eq:Susceptibility}. In the UV regime, we probe small distances within the system, and essentially the correlations behave as if the system were infinitely extended. By considering a hard-core interaction, we exclude a small region of real space, smaller than the extent of the system, within which correlations are zero. This further modifies the behavior. In the IR regime, we again obtain a saturation plateau, but in the UV regime, now we are probing distances smaller than the hard-core threshold, where the correlations in real space are zero, and so the ones in momentum space also exponentially fall to zero. Between the constant saturated value of correlations and the exponential drop-off, we transiently obtain the usual critical power law. The window of momenta within which we obtain this value is subject to variation depending on the extent of the system that is connected to the size of the colliding nuclei. Due to the fact that the large momenta boundary is primarily determined by the small distance, hard-core threshold which is constant, then it remains fairly constant, in contrast with the small momenta boundary which is primarily determined by the extend of the system and decreases for an increasingly larger system. Last but not least, these findings may shed new light upon intermittency analysis in a twofold way. Firstly, the small window within which the present examination predicts the critical power law will reside, is inevitably transferred to a corresponding window of transverse momentum space partitions, or in other words, to corresponding momentum separations of appropriate magnitude, where the intermittency analysis will reproduce the values of quantities, such as the intermittency index, that signal criticality. On the other hand, the IR plateau can be used to observe the non monotonicity of susceptibility when the freeze-out conditions of the system vary around the critical point.

\begin{table}[H]
    \centering
    \begin{tabular}{|c|c|c|c|}
        \hline
        $r ~ \rm (fm)$ & $k^{HC}_{\rm left} ~ \rm (MeV)$ & $k^{HC}_{\rm right} ~ \rm (MeV)$ & $\Delta \gamma$ \\
        \hline
        3 
        & 15.44 
        & 38.37
        & 0.455 \\
        \hline
        5 
        & 9.713 
        & 31.35
        & 0.489 \\
        \hline
        7 
        & 7.092 
        & 29.59
        & 0.510 \\
        \hline
        10 
        & 5.052 
        & 27.93
        & 0.510 \\
        \hline
        13 
        & 3.927 
        & 28.75
        & 0.524 \\
        \hline
    \end{tabular}
    \caption{Values of Quantities Defining the $-(2-a)$ Power Law Momenta Window}
    \label{tab:window_values}
\end{table}


\FloatBarrier

\end{document}